\documentclass[prd,11pt,aps,floats,nofootinbib,tightenlines,showpacs]{revtex4-1}
\usepackage{epsfig,graphicx}
\usepackage{psfrag}
\usepackage{color}
\usepackage{amsmath}
\usepackage{amsfonts}
\usepackage{amssymb}
\usepackage{textcomp}
\usepackage{multirow}
\usepackage{subfigure}
\usepackage[breaklinks=true,colorlinks=true,linkcolor=blue,urlcolor=blue,citecolor=blue]{hyperref}
\usepackage[normalem]{ulem}

\newcommand{\mcomment}[1]{}

\graphicspath{{Plots/}}

\begin{document}
	
	
	\title {Susceptibilities of conserved charges of hadronic matter}
	\author{Somenath Pal}
	\address{Variable Energy Cyclotron Centre, 1/AF, Bidhan Nagar , Kolkata-700064, India}
		\email{s.pal@vecc.gov.in}
	
	
	\def\be{\begin{equation}}
		\def\ee{\end{equation}}
	\def\bearr{\begin{eqnarray}}
		\def\eearr{\end{eqnarray}}

	\begin{abstract}
	An effective description of hadronic matter in terms of "effective free particles" is presented. Repulsive interactions among the hadrons have been modelled by Excluded volume correction by considering the hadrons as liquid drops, whose volume is inversely proportional to the cube root of the mass of the species. The only adjustable parameter in the model is pion radius. We get better agreement to the lattice results as compared to the previous studies.
	\end{abstract}
	\maketitle
	
	\section{Introduction}
An infinitely large hadronic system in thermodynamical equilibrium is specified by the temperature ($T$) and chemical potential ($\mu$). The total energy density determined by the $T$ and $\mu$ depends on the particle content of the system and their interactions. However, for a particular temperature, specification of chemical potential does not uniqely fix the particle content because the chemical potential in Grand Canonical Ensemble does not impose the conservation of the number of individual particle species. The particle content can be varied at a particular temperature and chemical potential by suitably choosing the net and total number of particles with infinite number of possibilities. The infinite number of possible configurations of the degrees of freedom at a particular temperature and chemical potential raises the need for specification of the interactions among the particles, which, for strongly interacting matter, is not known exactly. Effective models have been used to incorporate interactions among the particles for several decades in an attempt for the suitable description of the interactions prevailing in the system with partial success. The drawback of these models is that these models contain adjustable parameters, values of which are unknown. This hinders a proper description of the hadronic system in the effective model approach. To describe the interactions in a hadronic system without any adjustable parameter, we must formulate the problem in terms of effective free noninteracting particles without the presence of any explicit interaction term. Hadron Resonance Gas (HRG) model~\cite{Dashen:1969ep} takes the attractive interactions among the hadrons into account without any additional adjustable parameter. In spite of its tremendous success, HRG model fails to reproduce the charge susceptibilities predicted by lattice QCD, implying some discrepancy in the quantification of the interactions among the hadrons. Some intrinsic properties of hadrons known from the experiments are the masses, the conserved charges and spin degeneracies. Among these properties, mass is equivalent to energy. Interactions among the particles can also be introduced by suitably modifying the energy levels of the free particles in the system. For an isolated hadronic particle, the difference between the mass of a hadron and the total mass of the valence quarks forming it can be linked to the interactions among the constituent quarks. On the other hand, when the quark-gluon deconfined system hadronizes, the strong interactions among the quarks bind them to form various hadronic particles collectively having masses larger that the total mass of the constituent quarks inside all the hadrons. This mass difference carries the information regarding the interactions prevailing among the constituent quarks of a hadronic system as a whole. Quark-gluon deconfined state makes transition to the hadronic state only when the mass differences carrying the information about the interactions among the valence quarks inside the hadrons, to which the transition is happening, is compatible with the interactions among the different hadron species in the hadronic state. This indicates that it should be possible to describe the particle content of the hadronic matter in terms of mass, conserved charges, spin degrees of freedom and some other intrinsic properties of the hadrons, without any explicit reference to the interactions among the hadronic particles. By modifying the number densities of each particle species, one can control the interactions among the hadrons . We therefore attempt in this paper to take the hadronic interactions into account by means of effective abundance of different hadronic species treated as free particles and investigate how far the lattice results be reproduced.
	
Since energy is an additive quantity, it is therefore logical to start from the expression of energy density ($\epsilon$) as a function of these intrinsic properties, when one attempts to find the thermodynamic quantities of the hadronic matter.

	\begin{equation}
		\epsilon = \Sigma_i \tilde{\epsilon}_i  = \Sigma_i \frac{g_i}{2\pi^2}\int_0^{\infty} \frac{p^2 \sqrt{p^2 + m_i^2}}{e^{(\sqrt{p^2 + m_i^2} - \mu_i + \nu_i(T,\mu_B,\mu_Q,\mu_S,\{\mathcal{G}_i\}))/T} \pm 1}dp
		\label{total_energy_density}
	\end{equation}
	
	where, $\mu_i = B_i \mu_B + Q_i \mu_Q + S_i \mu_S$ is the part controlling the net conserved charges related to the grand potential and $\nu_i(T,\mu_B,\mu_Q,\mu_S,\{\mathcal{G}_i\})$ is the shift in the energy level of the ith hadron species due to interactions depending on all the intrinsic properties $\{\mathcal{G}_i\}$. Using the excluded volume model~\cite{Dashen:1969ep,Rischke:1991ke,Begun:2012rf}, $\nu_i(T,\mu_B,\mu_Q,\mu_S,\{\mathcal{G}_i\}))=P v_i$, where, $v_i=\frac{16}{3}\pi( r_i+r'(T,\mu_B,\mu_Q,\mu_S))$ is the excluded volume of hadronic species $i$; $r_i=\sqrt{\frac{3}{5}}(\frac{m_{\pi}}{m_i})^{1/3} r_{\pi}$ is the mechanical radius of hadronic species $i$, inspired from the liquid drop model~\cite{polyakov}; $r'(T,\mu_B,\mu_Q,\mu_S))=\frac{\Sigma r_i n_i^{1/3}}{\Sigma n_i^{1/3}}$ is the effective readius of all the other particles in the system combined as seen by any one particle. 
	
	We want to find an alternative picture of the same system in which all the interactions among the particles can be effectively taken into account by changing the equilibrium number densities of the free particle species. In this alternative picture of the system, the particles interact only via elastic collisions. From kinetic theory, the expression for number density of noninteracting particles in equilibrium with the condition of momentum conservation in any elastic collision happening in the system is given by~\cite{groot1980relativistic}
	
	\begin{equation}
		n = \Sigma_i n_i = \Sigma_i \frac{g_i}{2\pi^2} m_i^2 T \mathcal{K}_2 \Big (\frac{m_i}{T}\Big ) e^{\frac{\mu_i  \mathcal{E}}{T}}
		\label{number_density}
	\end{equation}
	Here, $g_i$ is the spin degeneracy of hadronic species $i$, $m_i$ is the mass, $\mathcal{K}_1$ and $\mathcal{K}_2$ are modified Bessel functions.  We have considered a common shift factor $\mathcal{E}$ in the effective chemical potential so that the effective degrees of freedom are populated with the same weightage on their mass. $\{\mathcal{G}_i\}$ represents all the intrinsic properties of the hadrons. The above formula is identical to the formula for the number density with Maxwell-Boltzmann distribution because the process of applying momentum conservation to any two-particle collision inside a system with different particle species makes the particles distinguishable, and hence, the effective statistics changes. The situation would not be the same for a system of particles of a single species. The distribution function of each species gets modified due to inter-species collisions which comes out to be the MB statistics explicitly. Now, from eq.~\ref{number_density}, we write the pressure $(P)$  of a system of free particles:
	\begin{equation}
		P = \Sigma_i \int n_i d\mu_i = \Sigma_i P_i = \frac{1}{\mathcal{E}}\Sigma_i \frac{g_i}{2\pi^2} m_i^2 T^2 \mathcal{K}_2 \Big (\frac{m_i}{T}\Big ) e^{\frac{\mu_i  \mathcal{E}}{T}}
	\end{equation}
The energy density is given by
	\begin{eqnarray}
		\epsilon &=& \Sigma_{i} \epsilon_i = -\frac{1}{T}\frac{\partial P_i}{\partial(1/T)} 
		 \Big |_{\frac{\{\mu_i\}}{T}}\nonumber \\	&=&\frac{1}{\mathcal{E}} \Sigma_i g_i \frac{m_i^4}{2 \pi^2} \Big [3 \frac{T^2}{m_i^2} \mathcal{K}_2 \Big (\frac{m_i}{T} \Big ) + \frac{T}{m_i} \mathcal{K}_1 \Big (\frac{m_i}{T} \Big ) \Big ] e^{\frac{\mu_i  \mathcal{E}}{T}}
		\label{total_energy_density_1}
	\end{eqnarray}
Comparing equations~(\ref{total_energy_density}) and ~(\ref{total_energy_density_1}), we can calculate $\mathcal{E}(T,\mu_B,\mu_Q,\mu_S,\{\mathcal{G}_i\})$ numerically. It is important to mention that no attempt will be made to find the analytic expression for $\mathcal{E}(T,\mu_B,\mu_Q,\mu_S,\{\mathcal{G}_i\})$ in this work.
\section{Results}
\begin{figure}[h!]
    \begin{tabular}{c|c}
       \includegraphics[scale=0.45]{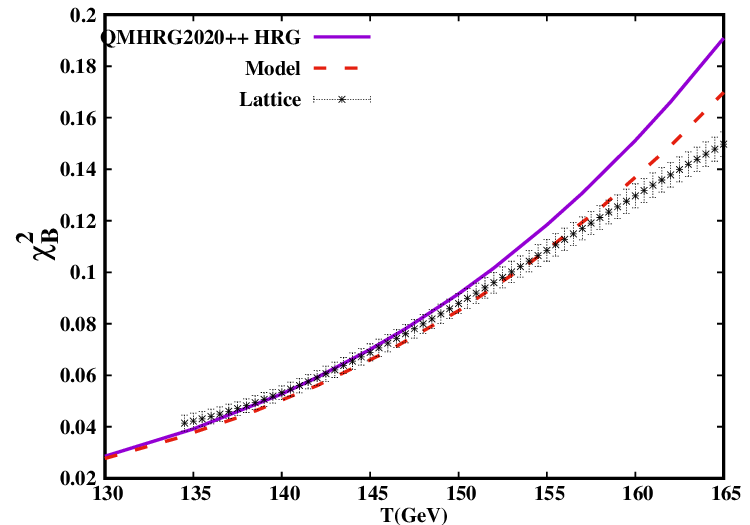}&
       \includegraphics[scale=0.45]{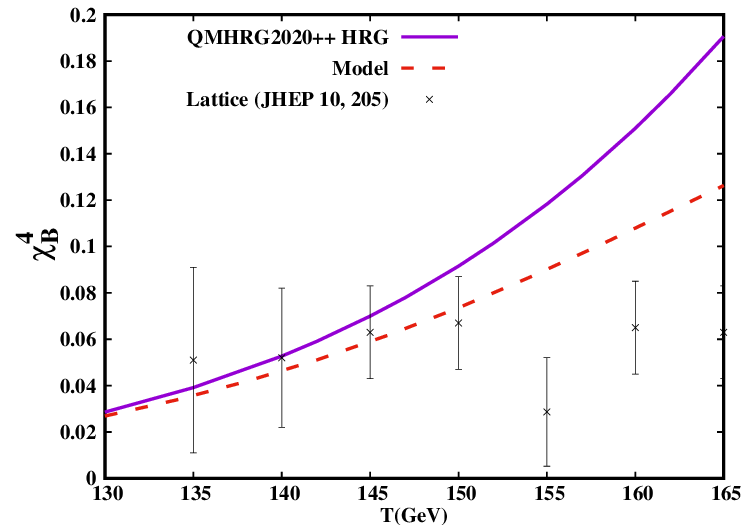}\\
       \includegraphics[scale=0.45]{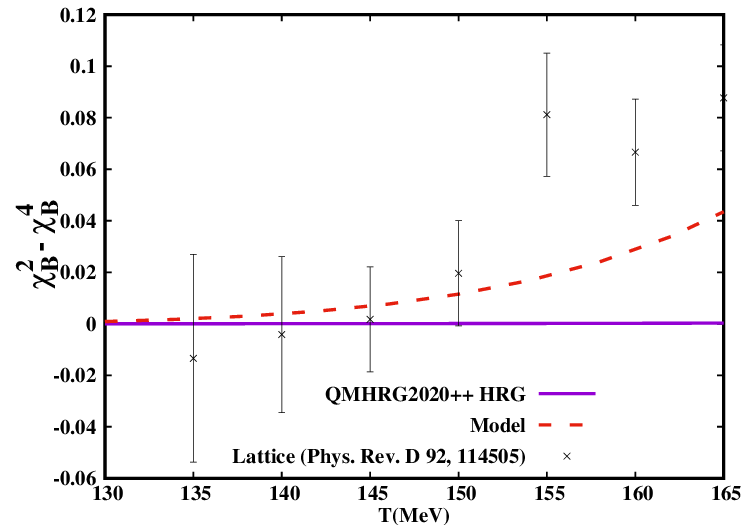}&
       \includegraphics[scale=0.45]{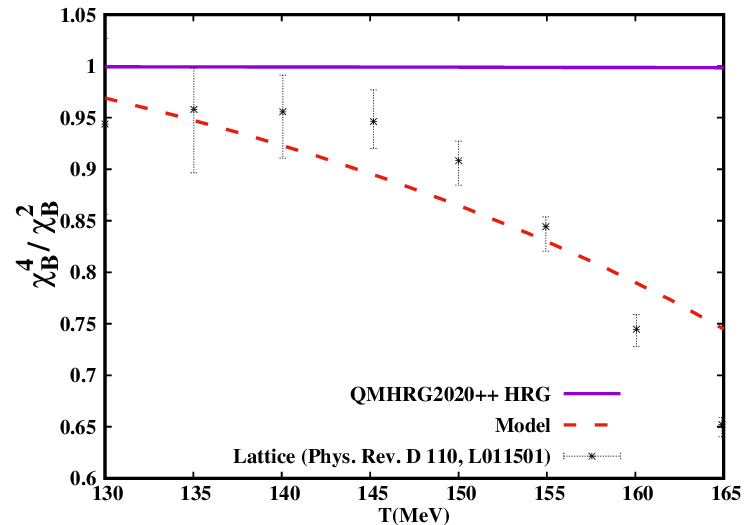}\\
       \includegraphics[scale=0.45]{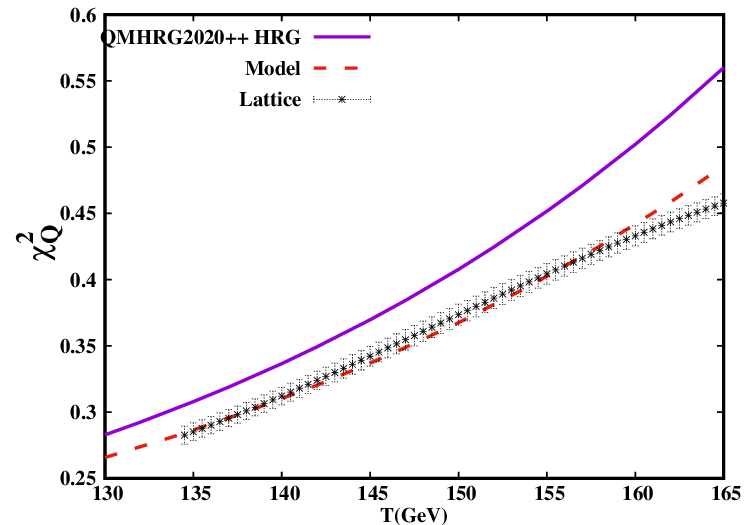}&
       \includegraphics[scale=0.45]{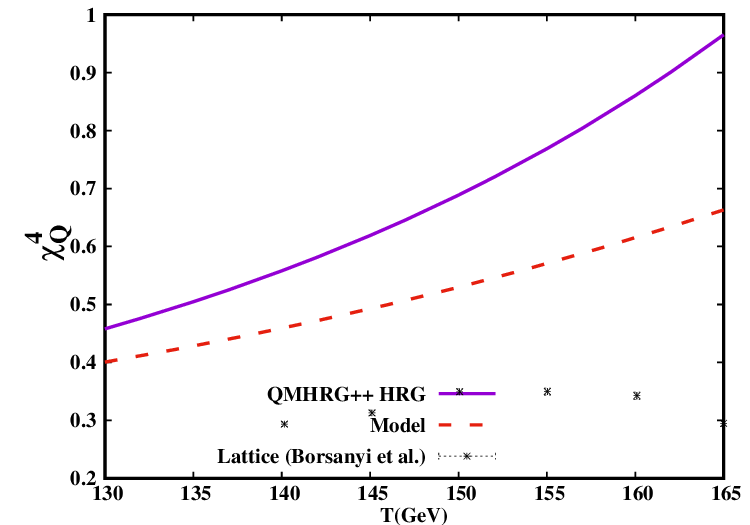}\\
       \includegraphics[scale=0.45]{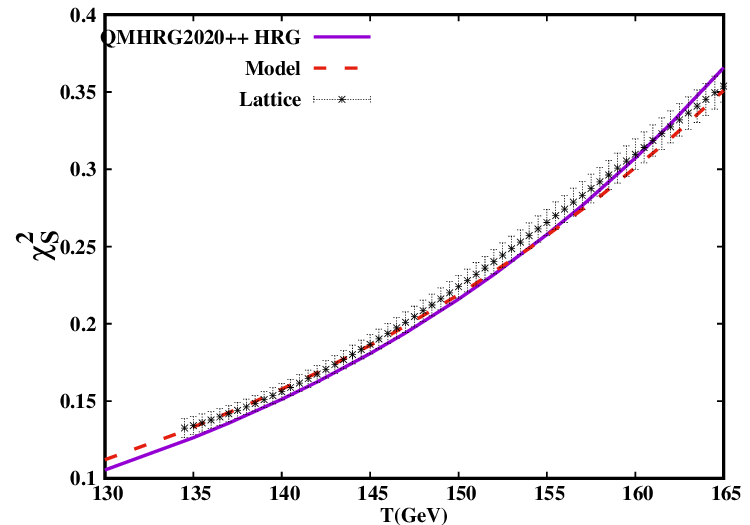}&
       \includegraphics[scale=0.45]{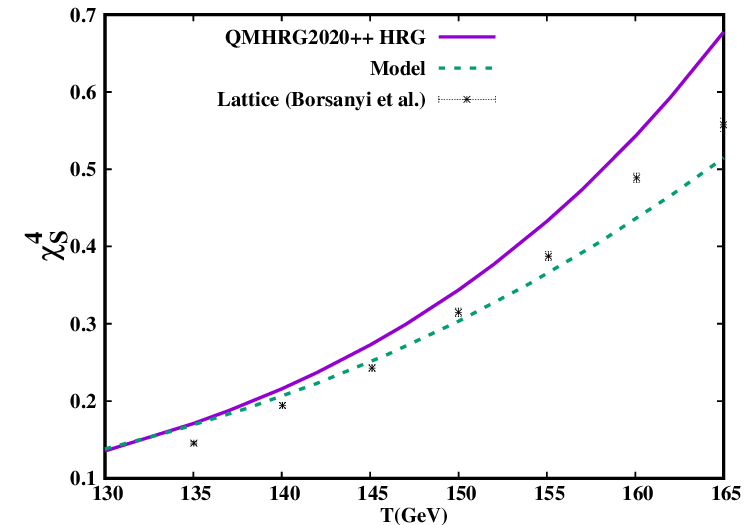}\\
       \includegraphics[scale=0.45]{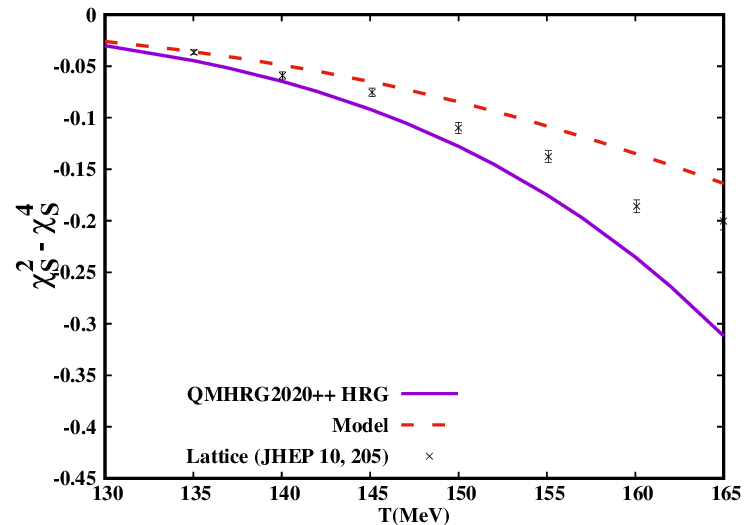}&
       \includegraphics[scale=0.45]{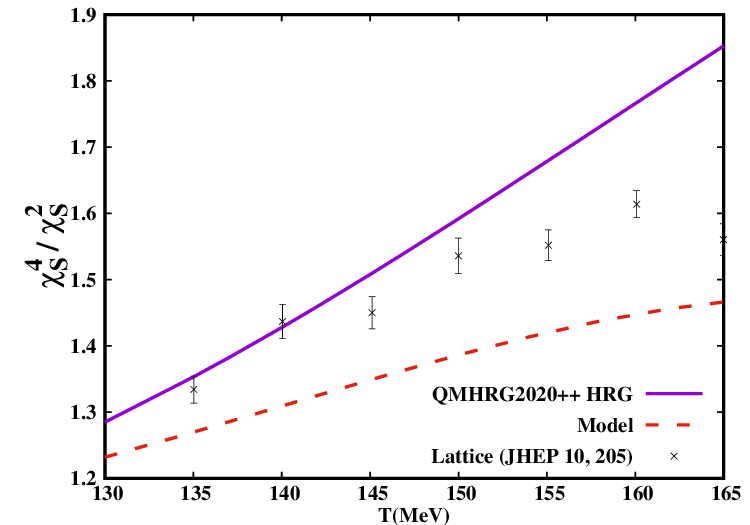}\\
    \end{tabular}
\end{figure}
\begin{figure}[h]
	\begin{tabular}{c|c}
\includegraphics[scale=0.45]{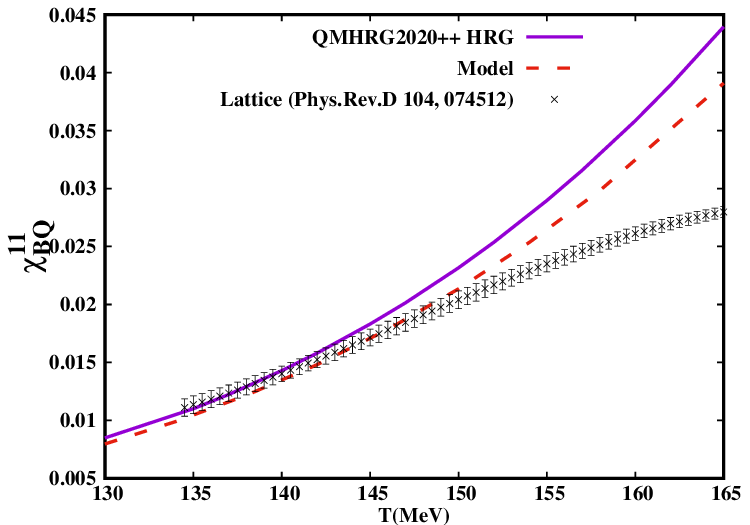}&
\includegraphics[scale=0.45]{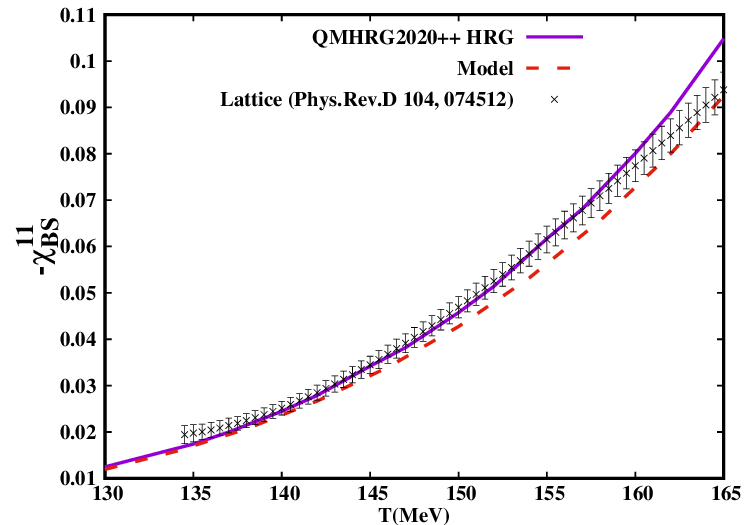}
\end{tabular}
\caption{Susceptibilities of conserved charges at zero chemical potentials.}
\label{pion_1}
\end{figure}
We take pion radius $r_{\pi}$ = 0.32 $fm$ and investigate the conserved charge susceptibilities at zero chemical potentials. The list of considered hadrons has been named QMHRG2020++, which includes some additional hadrons as compared to the QMHRG2020 list.It can be seen that the baryon and electric charge susceptibilities are well reproduced, except $\chi_Q^4$, lattice calculations of which suffers from serious numerical artifact. Agreement in the higher order strange sector is also not good which may be due to the undiscovered high mass strange hadrons.

\section{Summary}
It is expected that this simple model is not able to describe all the susceptibilities because the calculation of excluded volumes of the hadronic species based on the spherical liquid drop model is an oversimplified treatment. Discovery of more higher mass particles in the future will help us improve the results.

	 \section{Acknowledgment}
	SP acknowledges financial support from the Department of Atomic Energy, India. Tha author thanks Amaresh Jaiswal, Jishnu Goswami, Hiranmaya Mishra and Saumen Dutta for useful discussions.

	\bibliography{paper}
	
\end{document}